\documentclass[aps,prl,reprint,showpacs,floatfix,superscriptaddress,preprintnumbers]{revtex4-1}

\usepackage{graphicx}
\usepackage{color}
\usepackage{enumerate}
\usepackage{amssymb}
\usepackage{amsmath}
\definecolor{link_blue}{RGB}{45,48,146}
\usepackage[colorlinks,urlcolor=link_blue,citecolor=link_blue,linkcolor=link_blue,pdfstartview=FitH]{hyperref}

\usepackage{bibunits}

\usepackage[utf8]{inputenc}
\usepackage[T1]{fontenc}
\usepackage{txfonts} 

\begin{document}

\title{Vortex Thermometry for Turbulent Two-Dimensional Fluids}

\author{Andrew J. Groszek}
\affiliation{School of Physics and Astronomy, Monash University, Victoria 3800, Australia}

\author{Matthew J. Davis}
\affiliation{School of Mathematics and Physics, University of Queensland, Queensland 4072, Australia}

\author{David M. Paganin}
\affiliation{School of Physics and Astronomy, Monash University, Victoria 3800, Australia}

\author{Kristian Helmerson}
\affiliation{School of Physics and Astronomy, Monash University, Victoria 3800, Australia}

\author{Tapio P. Simula}
\affiliation{School of Physics and Astronomy, Monash University, Victoria 3800, Australia}

\begin{abstract}

We introduce a new method of statistical analysis to characterise the dynamics of turbulent fluids in two dimensions. We establish that, in equilibrium, the vortex distributions can be uniquely connected to the temperature of the vortex gas, and apply this vortex thermometry to characterise simulations of decaying superfluid turbulence. We confirm the hypothesis of vortex evaporative heating leading to Onsager vortices  proposed in \href{https://doi.org/10.1103/PhysRevLett.113.165302}{Phys.~Rev.~Lett.~\textbf{113}, 165302 (2014)}, and find previously unidentified vortex power-law distributions that emerge from the dynamics.

\end{abstract}

\preprint{DOI: \href{http://doi.org/10.1103/PhysRevLett.120.034504}{10.1103/PhysRevLett.120.034504}}

\maketitle

\begin{bibunit}

Turbulence arises in chaotic dynamical systems across all scales, from mammalian cardiovascular systems, to climate, and even to the formation of stars and galaxies \cite{davidson_turbulence:_2015}. The unpredictability inherent to turbulent systems is further confounded by physical properties such as boundaries and spatial dimensionality, and due to its complexity, there is currently no unified theoretical framework to explain turbulence. As such, there is a need to develop new methods to characterise the evolution of turbulent states in order to provide further insights into this important problem.

Onsager developed a model of statistical hydrodynamics to describe turbulence in two-dimensional (2D) flows \cite{onsager_statistical_1949}. In this representation the fluid is modelled by an $N$-particle gas of interacting point-like vortices which can be characterised by an equilibrium temperature. As the bounded system of vortices has a finite configuration space, the entropy $S$ of the system has a maximum, and hence there is a range of energy $E$ where the absolute Boltzmann temperature $T = (\partial S / \partial E)^{-1}$ becomes negative \cite{onsager_statistical_1949, purcell_nuclear_1951, ramsey_thermodynamics_1956}. These states correspond to large-scale rotational flows known as Onsager vortices \cite{onsager_statistical_1949}.

In driven 2D incompressible fluids, negative temperature Onsager states are known to emerge dynamically, effectively giving rise to order from chaos. This peculiar phenomenon is understood to be associated with an inverse energy cascade, in which energy flows towards the largest length scales in the system \cite{kraichnan_inertial_1967, kraichnan_two-dimensional_1980}. However, it is not clear whether this process should carry over to the regime of superfluid turbulence due to the compressibility of the superflow \cite{numasato_direct_2010}. Despite this, many recent theoretical works involving superfluid Bose--Einstein condensates (BECs) have suggested that large-scale, same-sign Onsager vortex clusters play an important role in 2D quantum turbulence \cite{bradley_energy_2012, white_creation_2012, reeves_inverse_2013, neely_characteristics_2013, billam_onsagerkraichnan_2014, simula_emergence_2014, groszek_onsager_2016, skaugen_vortex_2016, skaugen_origin_2017, salman_long-range_2016}. In superfluids, much of the focus to date has been on decaying---rather than driven---turbulence~\cite{walmsley_quantum_2008, white_creation_2012, neely_characteristics_2013, rooney_persistent-current_2013, billam_onsagerkraichnan_2014, simula_emergence_2014, kwon_relaxation_2014, cidrim_controlled_2016, groszek_onsager_2016, salman_long-range_2016, navon_emergence_2016, seo_observation_2017, walmsley_coexistence_2017}, since this scenario removes the complications of stirring. A variety of concepts have been applied to the problem, including holographic duality \cite{chesler_holographic_2013, du_holographic_2015} and nonthermal fixed points \cite{nowak_superfluid_2011, nowak_nonthermal_2012, karl_strongly_2017}.

Experimentally, BECs provide unprecedented opportunities to investigate 2D superfluid turbulence due to the high degree of controllability available in these systems. It is now possible to create and image complex vortex configurations such as dipoles \cite{freilich_real-time_2010, neely_observation_2010, middelkamp_guiding-center_2011, kwon_periodic_2015}, few-vortex clusters \cite{navarro_dynamics_2013} and quantum von K{\'a}rm{\'a}n vortex streets \cite{kwon_observation_2016}. Many experiments have also been devoted to the study of quantum turbulence in both two- \cite{neely_characteristics_2013, rooney_persistent-current_2013, kwon_relaxation_2014, seo_observation_2017} and three-dimensional \cite{henn_emergence_2009, tavares_chaotic_2016, navon_emergence_2016} geometries. However, the formation of Onsager vortex structures in statistical equilibrium has not yet been reported. Recent theoretical works have suggested that one significant obstacle is the harmonic trapping commonly used in experiments, as vortex clusters appear to be suppressed in this geometry \cite{kwon_relaxation_2014, stagg_generation_2015, groszek_onsager_2016}. In addition, the detection of vortex circulation signs is experimentally difficult, and only recently have techniques been proposed \cite{powis_vortex_2014} and implemented \cite{seo_observation_2017} to achieve this. Analysis of turbulent dynamics is made even more challenging by current condensate imaging methods, which only allow a small number of frames to be captured for a single experimental realisation \cite{wilson_situ_2015}. As such, it is desirable to be able to characterise the state of a turbulent fluid using a robust method of statistical analysis that links the instantaneous microscopic configuration of the system to its macroscopic behaviour. Onsager's thermodynamical description of turbulence is one such method, and hence we propose to use its central observable---the vortex temperature---for this purpose. In contrast to velocimetry-based observables that require the measurement of the velocities of the atoms, the thermometry presented here only requires the measurement of the positions and circulation signs of the quantised vortices.

We first outline our method for measuring the temperature of the vortex gas, before examining a specific case of decaying superfluid turbulence using mean-field Gross--Pitaevskii simulations. In the dynamics, we observe that the vortex gas undergoes rapid equilibration before settling into a quasi-equilibrium state where it continues to heat adiabatically via vortex evaporation \cite{simula_emergence_2014}. We have discovered that in this evolution, the numbers of clusters, dipoles and free vortices follow robust power-laws with respect to the total vortex number. The existence of this quasi-equilibrium allows quantitative thermometry of the turbulent fluid.

%
%

\begin{figure}[t]
\includegraphics[width=0.95\columnwidth]{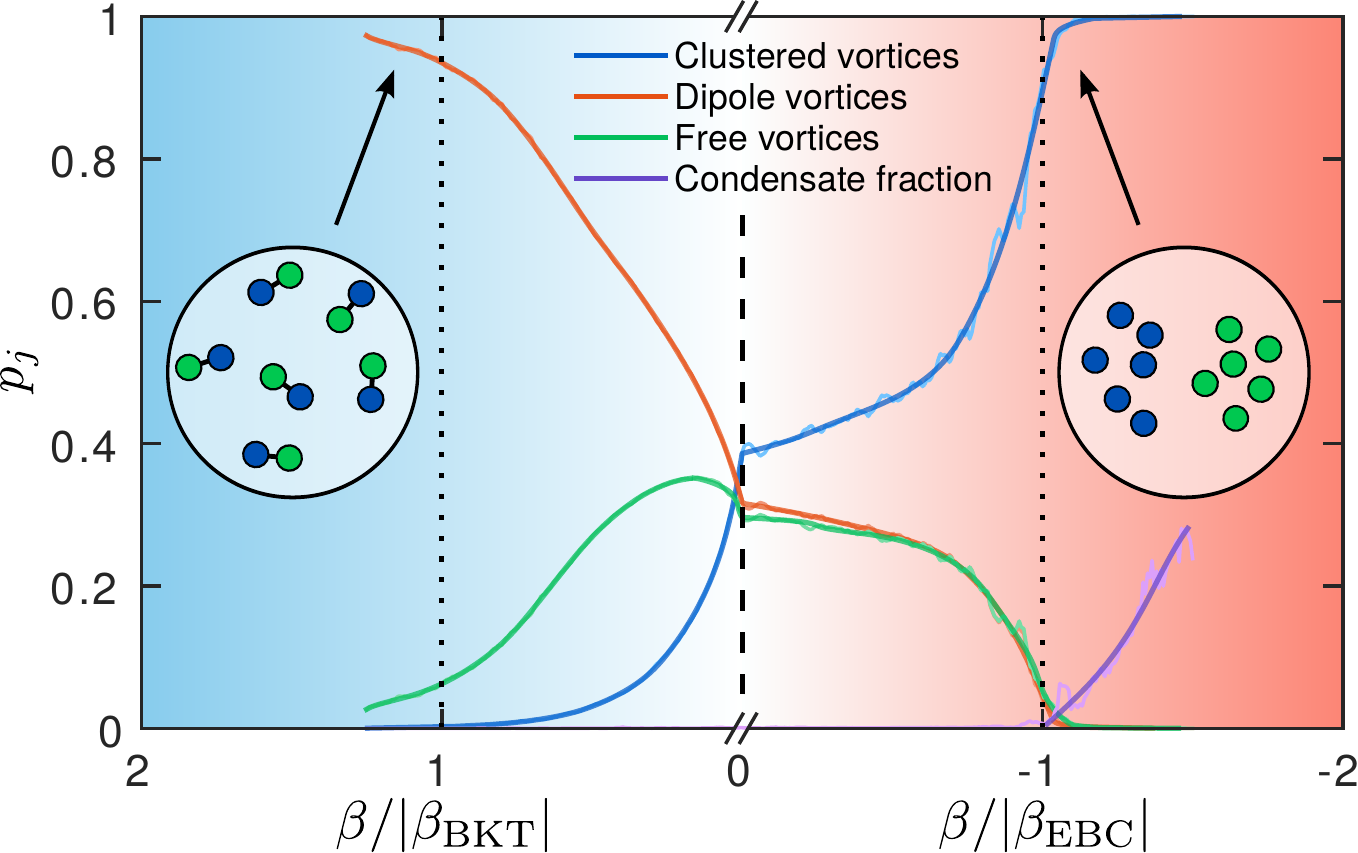}
\caption{\label{fig:MCdata} Fractional population $p_j=N_j/N_v$ of each component of the vortex gas (where $j \in \lbrace c$: clustered vortices, $d$: dipole vortices, $f$: free vortices$\rbrace$) and Einstein--Bose condensate fraction (described in text), as functions of inverse temperature $\beta$. The fluctuating faint lines show the raw data, while the smooth dark lines show cubic spline fits. The negative temperature axis is scaled by the critical temperature $|\beta_{\rm EBC}|$, and the positive temperature axis by $|\beta_{\rm BKT}|$, causing an apparent discontinuity in the slopes at $\beta=0$. The dashed vertical line indicates $\beta=0$ and the dotted vertical lines highlight the two critical temperatures. The shaded background represents the energy of the system (red and blue correspond to `hot'/high energy and `cold'/low energy, respectively). Schematic vortex configurations at each temperature extreme are depicted in the insets, where vortices (antivortices) are shown as blue (green) filled circles.}
\end{figure}

To calibrate the vortex thermometer, we use Monte Carlo (MC) simulations to map out the equilibrium vortex configurations as a function of the inverse temperature $\beta = 1/k_B T$, where $k_B$ is Boltzmann's constant. We do this for a system of $N_v=50$ point vortices with equal numbers of clockwise and anticlockwise circulations. Other values of $N_v$ are considered in the Supplemental Material \cite{supplement}. We use a point vortex Hamiltonian corresponding to a uniform fluid within a circular boundary of radius $R$ \cite{pointin_statistical_1976, simula_emergence_2014}, and set a hard vortex core of radius $0.003\,R$ to prevent energy divergences. As we vary the temperature across both positive and negative regimes, we quantify the effect on the vortex configuration using a vortex classification algorithm \cite{reeves_inverse_2013,valani_condensation_2016}. The algorithm uniquely divides the vortex gas into three separate components: clusters of $\geq 2$ like-sign vortices, closely bound vortex--antivortex dipoles, and relatively isolated free vortices (for further details, see Ref.~\cite{valani_condensation_2016}). We then calculate the number of clusters $N_c$, dipoles $N_d$ and free vortices $N_f$ as functions of temperature, and the resulting fractional population curves are presented in Fig.~\ref{fig:MCdata}.

At low positive absolute temperatures (left hand side of Fig.~\ref{fig:MCdata}), the vortex gas is at its `coldest', as both the energy and entropy are minimised. In this regime, bound vortex--antivortex dipole pairs dominate the configuration, as shown in the schematic inset of Fig.~\ref{fig:MCdata}. Above the Berezinskii--Kosterlitz--Thouless (BKT) critical temperature $\beta_{\rm BKT}$ \cite{berezinskii_destruction_1971, berezinskii_destruction_1972, kosterlitz_ordering_1973, hadzibabic_berezinskiikosterlitzthouless_2006, clade_observation_2009}, the vortex dipoles dissociate, causing an increase in both the energy and entropy. At $\beta=0$, the vortex configuration becomes a disordered arrangement of vortices and antivortices, thereby maximising the entropy. In the negative temperature region, low-entropy clusters of like-sign vortices tend to form (see schematic inset), and because of their high energy, these negative temperature states are `hotter' than those at positive temperature. Above the critical temperature $\beta_{\rm EBC}$, the vortices form an Einstein--Bose condensate (EBC), a state where the Onsager vortex clusters condense, as indicated by the saturation of the cluster population in Fig.~\ref{fig:MCdata} \cite{simula_emergence_2014, valani_condensation_2016, yu_theory_2016}. For a neutral vortex gas the two aforementioned critical temperatures are defined as $\beta_{\rm BKT} = 2/ E_{\circ}$ and $\beta_{\rm EBC} = -4/N_v E_{\circ}$ \cite{kraichnan_two-dimensional_1980}, respectively, where the energy $E_{\circ} = \rho_s \kappa^2 / 4 \pi$ is defined in terms of the superfluid density $\rho_s$ and the quantum of circulation $\kappa = h/m$, with $m$ being the mass of the condensed atoms.

Figure \ref{fig:MCdata} demonstrates that the dipole and cluster populations are monotonic functions of $\beta$---this is the key observation enabling thermometry of the vortex gas. Given an arbitrary vortex configuration in thermal equilibrium, we may determine its temperature by calculating the populations of clusters and/or dipoles and comparing the obtained values to the curves in Fig.~\ref{fig:MCdata}. Strictly, the $p_j(\beta)$ curves in the negative temperature region of Fig.~\ref{fig:MCdata} are dependent on the vortex number. However, we repeated our MC simulations for $N_v=100$ and $200$ vortices and found that, for the vortex numbers relevant to the dynamical simulations presented here, there is no qualitative change to the thermometry curves, and the quantitative change is not significant (see Supplemental Material~\cite{supplement}). The cluster and dipole fractions are not the only observables that vary monotonically with vortex temperature in our MC simulations. For example, both the energy and dipole moment of the vortex gas also fulfill this requirement~\cite{simula_emergence_2014}, and could therefore, in principle, be used for thermometry. However, of all variables considered, we have found that the cluster and dipole fractions provide the most robust thermometers.

Also shown in Fig.~\ref{fig:MCdata} is the Einstein--Bose condensate fraction, which quantifies the density of vortices in the largest cluster (for details, see Ref.~\cite{valani_condensation_2016}). For $\beta > \beta_{\rm EBC}$, the condensate fraction is zero, but when $\beta < \beta_{\rm EBC}$ it rises sharply. In this extreme temperature region, the other thermometers saturate and the condensate fraction becomes the relevant observable for vortex thermometry.

%
%

As an application of our vortex thermometer, we use it to characterise decaying turbulence in a disk-shaped BEC as previously studied in Refs.~\cite{simula_emergence_2014, groszek_onsager_2016}. We simulate the two-dimensional time-dependent Gross--Pitaevskii equation (GPE),
\begin{equation} \label{eq:GPE}
i \hbar \frac{\partial}{\partial t} \psi = \left[ - \frac{\hbar^2}{2m} \nabla^2 + V_{\rm tr} + g_{\rm 2D} \left| \psi \right|^2 \right] \psi,
\end{equation}
where $\psi \equiv \psi(\textbf{r},t)$ is the classical field of the Bose gas and $g_{\rm 2D}$ is the two-dimensional interaction parameter resulting from the $s$-wave atomic collisions. To obtain the uniform circular geometry, we use a two-dimensional power-law trapping potential of the form $V_{\rm tr} = \mu (r/R)^{50}$, where $r=\sqrt{x^2+y^2}$ is the radial distance from the axis of the trap, $\mu$ is the chemical potential, and $R \approx 171 \, \xi$ is the radius of the trap, measured in units of the healing length $\xi=\sqrt{\hbar^2/2 m \mu}$ \cite{groszek_onsager_2016}. The interaction parameter is set to $g_{\rm 2D} = 4.6 \times 10^4 \, \hbar^2 / m$. We solve the GPE using a fourth-order split-step Fourier method on a $1024^2$ numerical grid with a spacing of approximately $ \xi / 2$. Turbulence is generated by imprinting vortices into the phase of $\psi$ and evolving Eq.~\eqref{eq:GPE} for a short amount of imaginary time to establish the vortex core structures. We detect vortices and their circulation signs within $r < 0.98 \, R$ by locating singularities in the phase of the field.

\begin{figure}[t]
\includegraphics[width=\columnwidth]{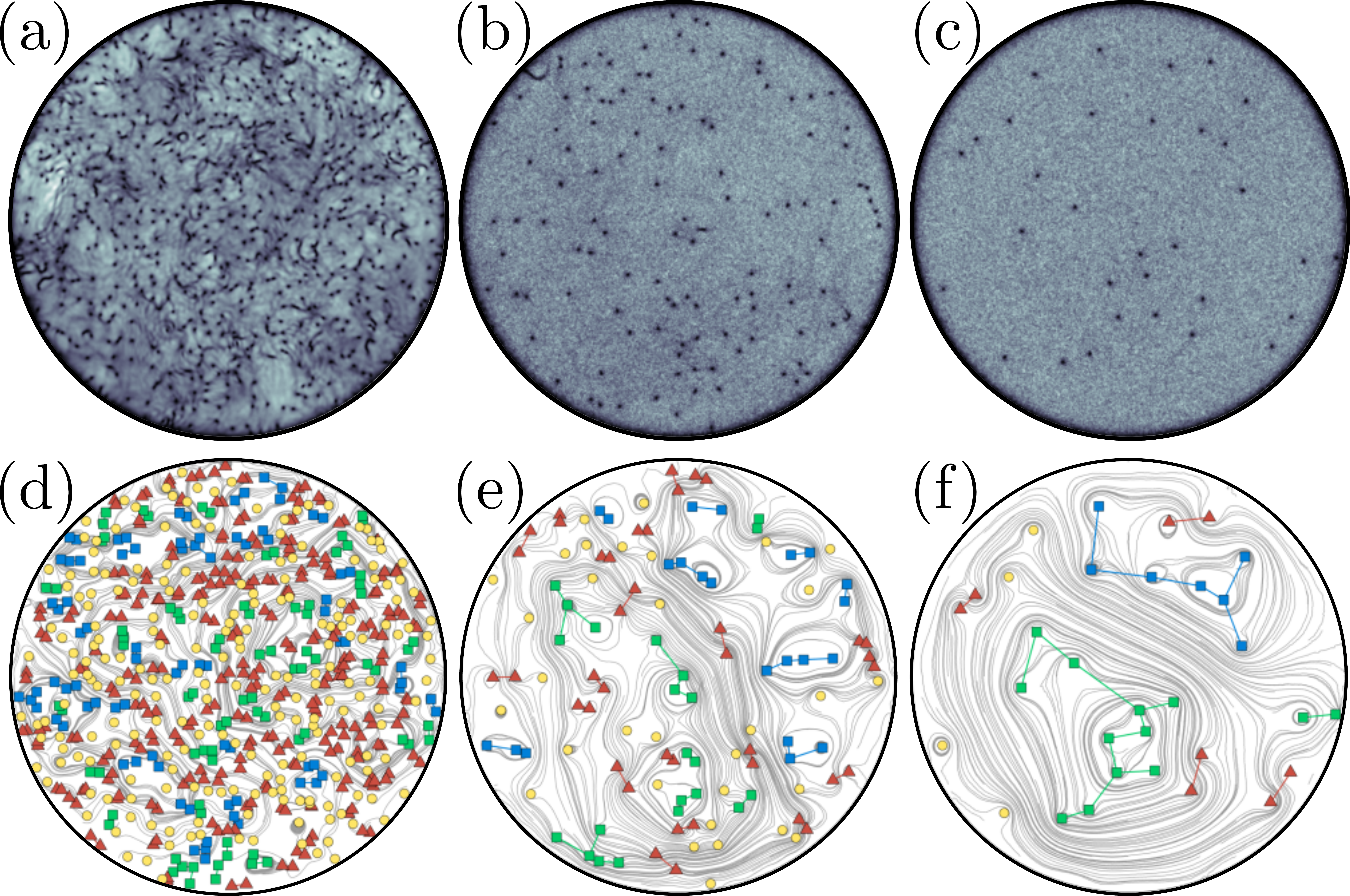}
\caption{\label{fig:GPE_screenshot} Freely decaying two-dimensional quantum turbulence. Panels (a)--(c) show the classical field density $|\psi|^2$, for respective times $t \approx (25,7500$ and $74 \, 000$) $\hbar/\mu$, and reveal the vortices as dark spots of zero density. The grayscale in each panel is normalised to the respective peak value of $|\psi|^2$. Panels (d)--(f) correspond to panels (a)--(c), respectively, and show the vortices in positive (negative) clusters as blue (green) squares, dipoles as red triangles, and free vortices as yellow circles. Note that each vortex dipole contains one vortex and one antivortex. The streamlines in (d)--(f) are obtained by calculating the incompressible component of the velocity field of the classical field describing the Bose gas.}
\end{figure}

The initial vortex configurations used in our GPE simulations are produced by randomly drawing $N_v=800$ vortex locations from a uniform distribution, with equal numbers of each circulation sign. The resulting state is well approximated to have $\beta \approx 0$, although the short imaginary time propagation step causes a small amount of cooling towards positive temperatures. As the turbulence decays, the vortices annihilate and the vortex gas evaporatively heats, resulting in the emergence of two large Onsager vortices at late times \cite{simula_emergence_2014, groszek_onsager_2016}. Three sample frames from a single simulation are shown in Fig.~\ref{fig:GPE_screenshot}, where panels (a)--(c) show the density $|\psi|^2$ of the fluid, and panels (d)--(f) show the corresponding vortex configuration after the vortex detection and classification steps. A Helmholtz decomposition \cite{bradley_energy_2012} has been used to extract the divergence-free component of the condensate velocity field, and the resulting streamlines are also shown in the lower panels. The Onsager vortex clusters are clearly visible in panel (f).

\begin{figure}[t]
\includegraphics[width=0.85\columnwidth]{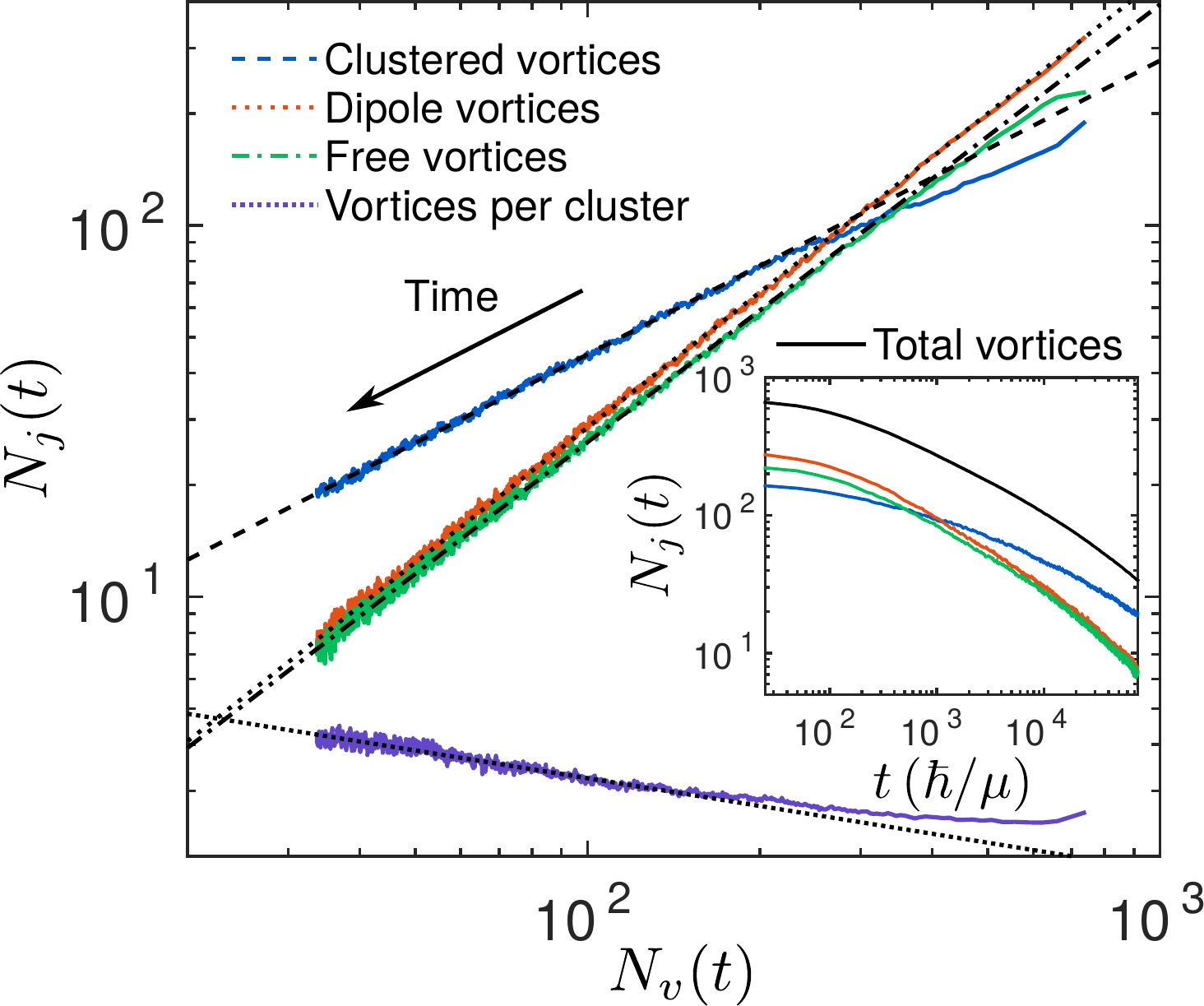}
\caption{\label{fig:number_decay} Decay of the vortex populations $N_j(t)$ (where $j \in \lbrace c$: clustered vortices, $d$: dipole vortices, $f$: free vortices$\rbrace$) in each component of the vortex gas, and the growth of the number of vortices per cluster $N_{vc}(t)$, as functions of the total vortex number $N_v(t)$. The data have been averaged over 80 simulations, with power-law fits shown as straight lines. Note that time flows from right to left in this figure. The inset shows the total number of vortices and the number of vortices in each component of the vortex gas as functions of time.}
\end{figure}

The number of clusters, dipoles and free vortices are shown in Fig.~\ref{fig:number_decay} as functions of both time $t$ (inset) and the total number of vortices $N_v(t)$. The time-dependent populations (inset) do not follow any simple function. However, the populations as functions of the total number of vortices (main frame) show clear power-law scaling behaviour. The corresponding power-laws are: $N_c \propto N_v^{\alpha}$, $N_d \propto N_v^{\gamma}$, $N_f \propto N_v^{\delta}$ and $N_{vc} \propto N_v^{\varepsilon}$, with measured values $\alpha = 0.79 $, $\gamma = 1.21$, $\delta = 1.18$, and $\varepsilon =-0.25$. These are suggestive of rational value power-laws with exponents $\alpha = 4/5$, $\gamma=\delta = 6/5$ and $\varepsilon=-1/4$. The mean number of vortices per cluster $N_{vc}\equiv N_c / N_{cl}$, where $N_{cl}$ is the total number of clusters of any size at a given time. To study the effects of system size on these power-laws, we have also considered two smaller disk-shaped systems of radii $R\approx49 \, \xi$ and $R\approx 85 \, \xi$ respectively, each with $N_v=100$ vortices initially imprinted. We find that the scaling behaviour is unchanged in these smaller systems, suggesting that the evolution of the vortex gas is underpinned by a universal microscopic process.

In this system, the primary cause of vortex number decay is the annihilation of vortex--antivortex dipoles. Despite this, the populations of dipoles and free vortices follow approximately the same power-law, demonstrating an interconversion between the vortex populations. However, a distinct power-law emerges for the vortex clusters. This behaviour points toward a two-fluid model, where the dipoles and free vortices behave as a weakly interacting thermal cloud, while the clusters act as a quasi-condensate whose relative weight grows over time as a result of vortex evaporative heating. Extrapolating the data toward $N_v\to0$ leads to the inevitable decay of all dipoles and free vortices, with only Onsager vortex clusters remaining. At this point, the rate of pair annihilation becomes insignificant in the dynamics due to the very low probability of vortex--antivortex collisions.

In Fig.~\ref{fig:number_decay}, the $N_d$ and $N_f$ curves are well described by the $N_v^{6/5}$ scaling throughout the dynamical evolution. The $N_c$ curve, on the other hand, only begins to follow the $ N_v^{4/5}$ power-law once the total vortex number has decayed to $N_v \lesssim 200$, suggesting that the statistical behaviour of the vortex gas changes at this point in the dynamics. In accordance with the existence of power-law scaling, we interpret this change to be the realisation of a state of quasi-equilibrium for the decaying turbulence. Under this quasi-equilibrium condition, the vortex evaporative heating process becomes adiabatic in the sense that the vortex gas is able to rearrange into a higher entropy configuration between the vortex annihilation events. For $N_v \gtrsim 200$, the vortex number decays too rapidly for this to be possible. This quasi-equilibrium condition is not a true equilibrium of the system, since vortex--antivortex annihilations and vortex--sound interactions are continuously driving energy from the vortices into the sound field. Presumably, the true equilibrium of the condensate will only be realised when all vortices have decayed and the total entropy of the system is maximised. In the Supplemental Material~\cite{supplement}, we present vortex number decay data for a range of other initial vortex configurations, observing in all cases evidence for the same power-law and quasi-equilibration behaviour.

%
%

\begin{figure}[t]
\includegraphics[width=0.9\columnwidth]{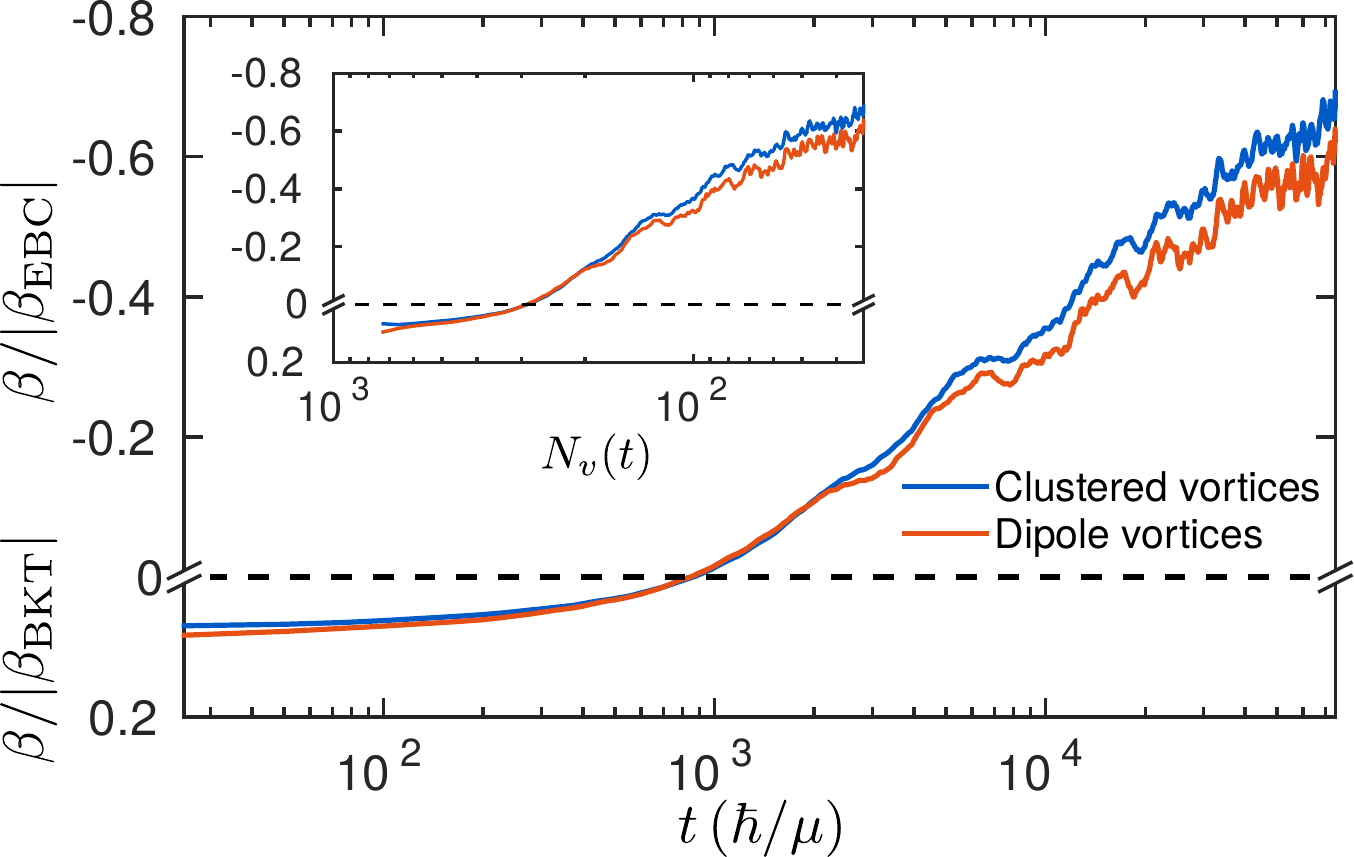}
\caption{\label{fig:dynamical_temperature} Inverse temperature of the vortex gas as a function of time, averaged over a set of dynamical GPE simulations. The temperature is measured independently using the populations of both clusters and dipoles. In the inset, the temperature readings from each thermometer is shown as a function of the total vortex number $N_v(t)$. As in Fig.~\ref{fig:MCdata}, the positive and negative temperature regions have been scaled by their respective critical temperatures, and a dashed horizontal line denotes $\beta=0$. The vertical axis of the inset is the same as for the main frame.}
\end{figure}

We now have an algorithm to assign a vortex temperature to the dynamical GPE simulations. We determine the fractional populations of vortex dipoles and clusters as a function of time, and use each of these to read off a temperature from the curves in Fig.~\ref{fig:MCdata}. The two resulting measurements of $\beta(t)$ are presented in the main frame of Fig.~\ref{fig:dynamical_temperature}. Both measurements show that the temperature of the vortex gas is spontaneously increasing as Onsager vortex clusters form, thereby confirming the evaporative heating scenario posited in Ref.~\cite{simula_emergence_2014}. At late times, a small discrepancy between the two temperature readings emerges, which we attribute to the compressibility of the fluid not accounted for in the MC model. The same temperature measurements are plotted as a function of the total vortex number in the inset. Based on our quasi-equilibrium interpretation discussed above, we note that the temperature reading is strictly only valid for $N_v \lesssim 200$ ($t \gtrsim 2000 \, \hbar / \mu$), since outside of this range the vortices are out of equilibrium and their temperature is not well defined. To obtain these curves, we have applied the thermometer calibrated with $N_v=50$ vortices, despite the fact that $N_v$ varies between $\approx 30$ and $\approx 200$ throughout the equilibrium dynamical evolution. In the Supplemental Material~\cite{supplement}, we show that using a thermometer calibrated with a different number of vortices does not affect the qualitative shape of the $\beta(t)$ curve.

%
%

We have developed a methodology that allows the temperature of point vortices in two-dimensional fluids to be determined using only the information about the vortex positions and their signs of circulation. We have applied the vortex gas thermometers to freely decaying two-dimensional quantum turbulent systems and quantitatively shown the transition to negative temperatures and the emergence of Onsager vortices due to the evaporative heating of the vortex gas \cite{simula_emergence_2014, groszek_onsager_2016}. Our vortex thermometers may also be useful for characterisation of turbulent classical fluids, as the continuous vorticity distributions can be approximated accurately by a discretised set of point vortices before performing the vortex classification and thermometry. This methodology may therefore open new pathways to quantitative studies of two-dimensional turbulence.

\begin{acknowledgments}
We thank Matthew Reeves for useful feedback on the manuscript. We acknowledge financial support from an Australian Government Research Training Program Scholarship (A.G.), the Australian Research Council via Discovery Projects No.~DP130102321 (T.S., K.H.), No.~DP170104180 (T.S.), No.~DP160103311 (M.D.) and the nVidia Research Grant Scheme. This research was undertaken with the assistance of resources from the National Computational Infrastructure (NCI), which is supported by the Australian Federal Government.
\end{acknowledgments}


%

\end{bibunit}

\begin{bibunit}

\pagebreak


\begin{center}
\large \textbf{Supplemental material for:\\
Vortex Thermometry for Turbulent Two-Dimensional Fluids}
\end{center}

\section{Initial condition dependence}

\begin{figure}[b!]
\includegraphics[width=\columnwidth]{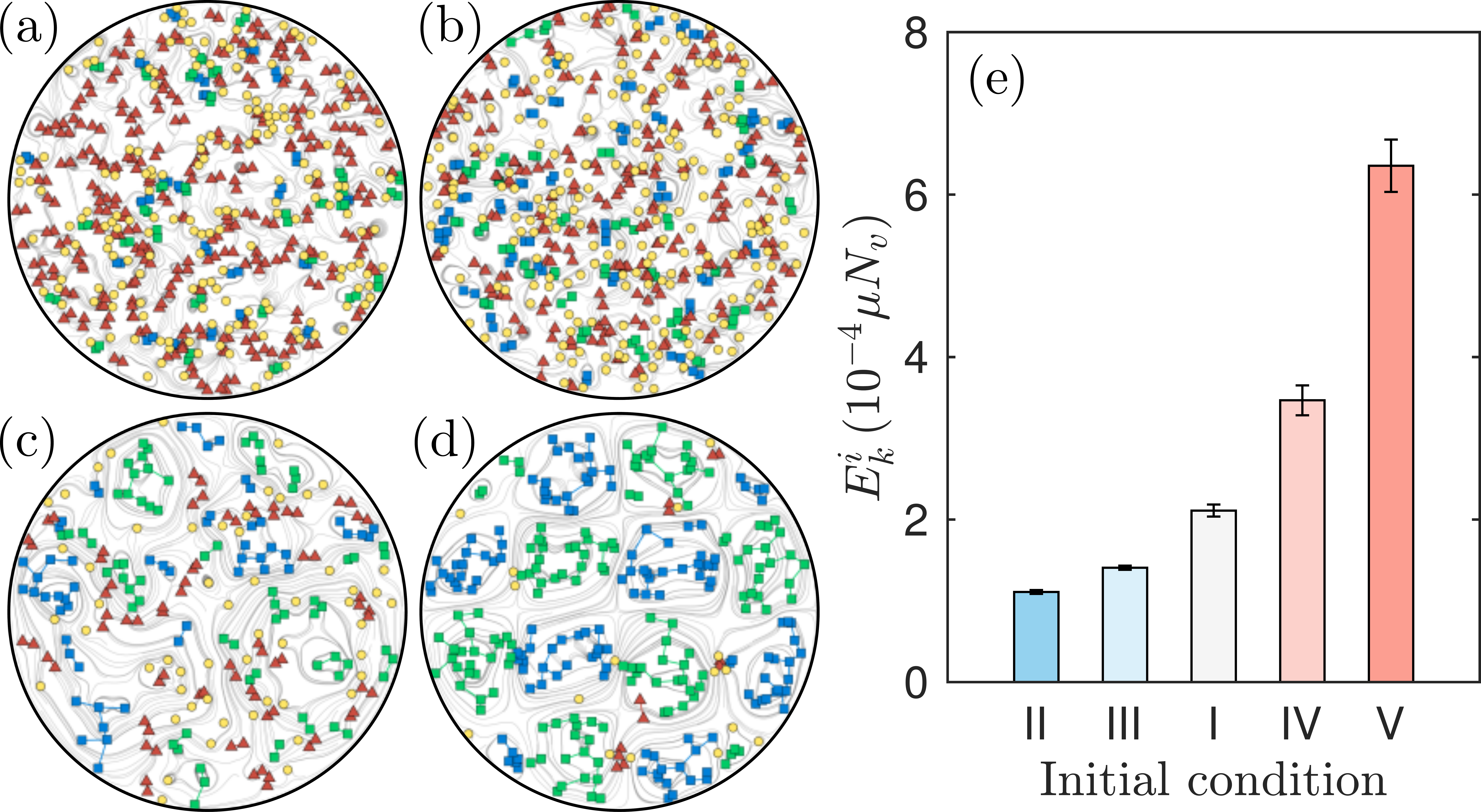}
\caption{\label{fig:initial_conditions} Examples of initial vortex configurations for our turbulent Gross--Pitaevskii simulations. Panels (a)--(d) correspond to cases II--V, respectively, and show the vortices in positive (negative) clusters as blue (green) squares, dipoles as red triangles, and free vortices as yellow circles. Note that each vortex dipole contains one vortex and one antivortex. The streamlines in each frame are obtained by calculating the incompressible component of the velocity field of the classical field describing the Bose gas. Panel (c) shows the dynamically stirred configuration immediately after the stirrer has been switched off. See also Fig.~2(d) in the main text, which shows the initial condition for case I. Panel (e) shows the incompressible kinetic energy $E^i_k$ per vortex for each of the five cases, averaged over 80 (10) initial conditions generated for case I (cases II--V). Error bars denote one standard deviation. The shading is indicative of how `cold' (blue) or `hot' (red) a given initial state is.}
\end{figure}

To assess the sensitivity of the observed power-laws (Fig.~3 of the main text) to the choice of initial vortex configuration, we have run Gross--Pitaevskii simulations with a diverse range of initial conditions. In addition to the randomly sampled initial condition (case I) described in the main text, we have considered four other types of initial state. The cases II and III are configurations with lower incompressible kinetic energy $E_k^i$, created by imprinting the vortices randomly throughout the condensate as dipole pairs with sizes $8 \, \xi$ and $12 \, \xi$, respectively (before the imaginary time propagation step). For case IV, motivated by experiments (e.g.~\cite{raman_evidence_1999, neely_observation_2010, kwon_relaxation_2014}), the vortex creation is simulated dynamically by stirring an initially unperturbed condensate with a repulsive Gaussian potential of waist $30 \, \xi$ and amplitude $5 \, \mu$. The stirring potential is moved back and forth with centroid position $x_\circ(t)=100 \xi \cos(2\pi \mu t / 1050 \hbar)$ for four periods, and then ramped down to zero over a fifth period. Finally, a large incompressible kinetic energy in case V is initiated by imprinting a periodic square array of vortex clusters with alternating circulation sign, each with a radius of $\approx 43 \, \xi$ and containing up to $25$ randomly placed vortices. Examples of initial conditions for cases II--V are shown in Fig.~\ref{fig:initial_conditions}(a)--(d), where the vortices have been classified into clusters, dipoles and free vortices as described in the main text. The corresponding mean incompressible kinetic energy for each initial condition is shown in Fig.~\ref{fig:initial_conditions}(e). This energy is defined $E_k^i = (m/2) \int |\psi|^2 |\mathbf{v}^i|^2 \mathrm{d}^2\textbf{r}$, where the $\mathbf{v}^i(\textbf{r})$ is the divergence-free component of the total velocity field.

\begin{figure}[b!]
\includegraphics[width=0.95\columnwidth]{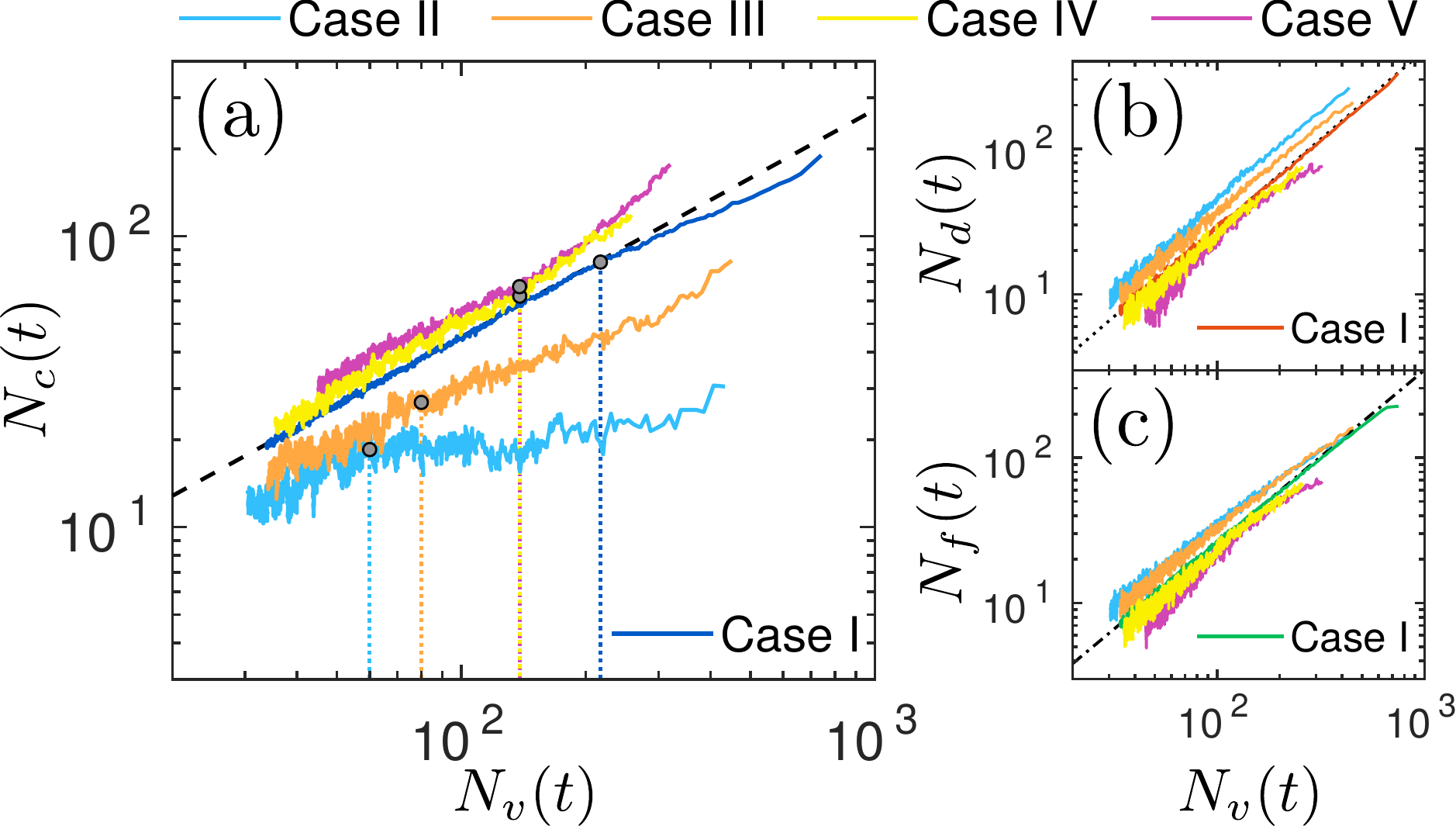}
\caption{\label{fig:initial_condition_dependence} Vortex decay curves for initial conditions I--V described in the text. The curve for case I (cases II--V) has been averaged over 80 (10) simulations. The three panels separately show (a) clusters, (b) dipoles, and (c) free vortices. Note that the curves for case I and their power-law fits are identical to those in Fig.~3 of the main text. In panel (a), dotted vertical lines and filled circles indicate the value of $N_v$ at which each cluster decay curve begins to approximate the $N_c \sim N_v^{4/5}$ power-law.}
\end{figure}

The resulting number decay curves for each vortex type are shown in Fig.~\ref{fig:initial_condition_dependence}(a)--(c). The dipole and free vortex decay curves [panels (b) and (c), respectively] remain relatively unchanged across different initial configurations. By contrast, the clusters [panel (a)] show clear variation across the set of initial conditions, suggesting that initially the system is in fact behaving very differently under each constraint. However, despite initial differences (at large $N_v$), all cluster decay curves eventually exhibit behaviour consistent with the power-laws obtained in Fig.~3 of the main text, demonstrating a loss of memory of the initial vortex configuration. This provides further evidence that these power-laws correspond to a state of quasi-equilibrium in which the vortex gas should have a well-defined temperature, as argued in the main text. In Fig.~\ref{fig:initial_condition_dependence}(a), the approximate value of $N_v$ at which the $N_c$ curve begins to follow the $N_v^{4/5}$ power-law is also highlighted, which we interpret as being the point at which the vortex gas reaches quasi-equilibrium.  

\begin{figure}[t]
\includegraphics[width=0.85\columnwidth]{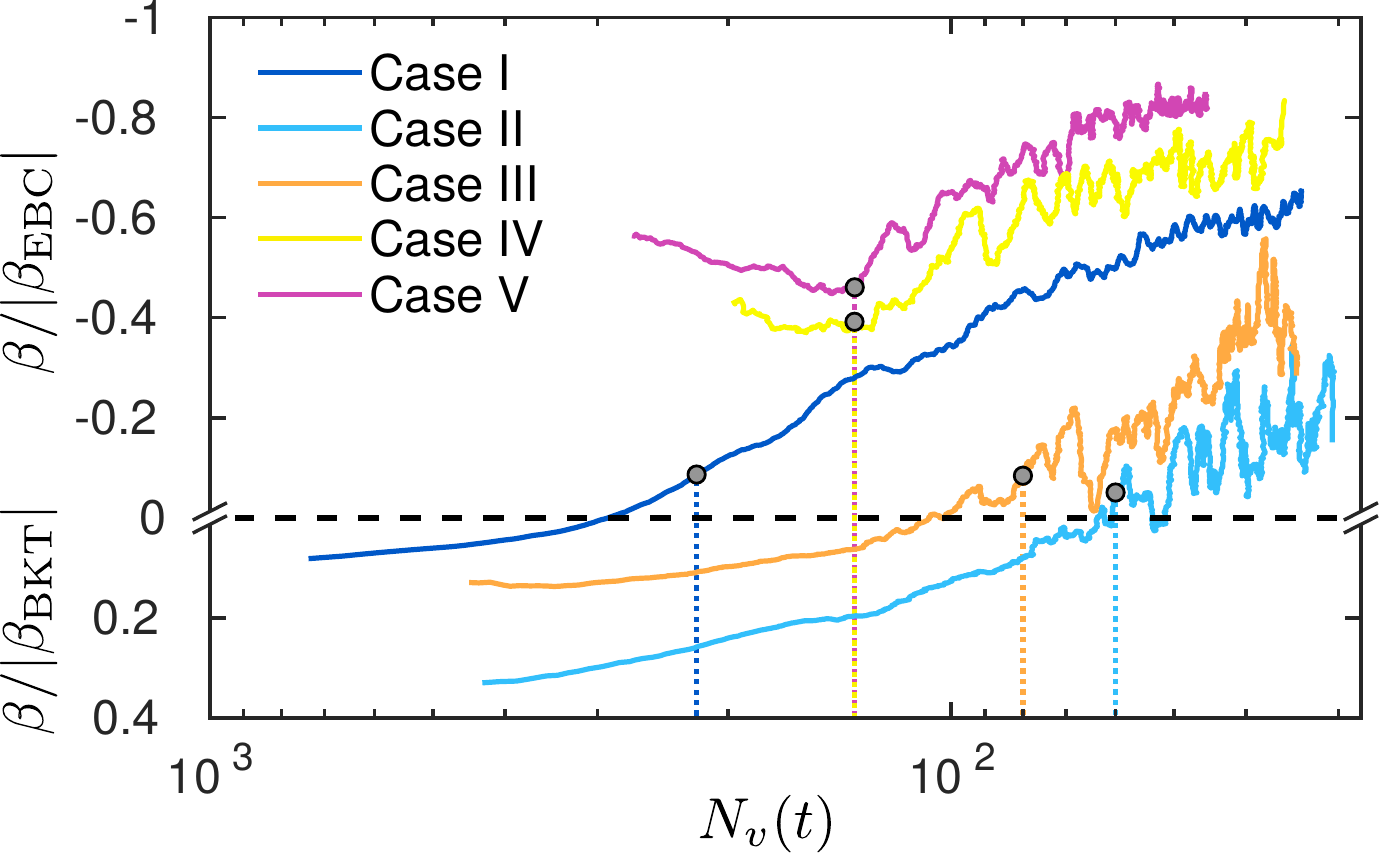} 
\caption{\label{fig:S2} Inverse temperature readings for all five initial conditions (as described in the text) as a function of the total vortex number $N_v(t)$. Each curve is ensemble averaged over multiple simulations, and is calculated from the mean of the cluster and dipole thermometry measurements. The vertical dotted lines and filled circles are from Fig.~\ref{fig:initial_condition_dependence}(a). As in Figs.~1 and 4 of the main text, the positive and negative temperature regions have been scaled by their respective critical temperatures, and a dashed horizontal line denotes $\beta=0$.}
\end{figure}

Applying our thermometer (Fig.~1 from main text) to all five initial conditions, we obtain the temperature readings for each, which are presented in Fig.~\ref{fig:S2}. Here, the temperature is calculated from the mean of the cluster and dipole thermometer measurements, which are themselves ensemble averaged over 80 (10) simulations for case I (cases II--V).
These curves show a clear dependence on initial condition, with the low energy configurations (cases II and III) being consistently colder than those with high energy (cases IV and V). The random initial configuration (case I) lies between the two extremes. The approximate value of $N_v$ at which the vortex gas appears to reach equilibrium in each case [see Fig.~\ref{fig:initial_condition_dependence}(a)] is also shown. Even before this point (i.e.~for larger $N_v$), the vortex thermometer provides a plausible temperature reading, but the measurement is not reliable if the vortex gas is out of equilibrium. In cases IV and V, the equilibration point corresponds to a turning point in the temperature curve, providing further evidence for our interpretation of the vortex gas equilibrium condition.

\section{Vortex number dependence of thermometry}

In Fig.~4 of the main text, the temperature measurement of the decaying turbulence for case I was obtained using a single thermometer calibrated with $N_v=50$ vortices. Strictly, this thermometer is only quantitatively valid for the short time in the dynamical evolution when $N_v \approx 50$, and additional thermometers should be calibrated to obtain more accurate measurements for other vortex numbers. Here we demonstrate that changing $N_v$ in the Monte Carlo simulations has only a small effect on the thermometry curves, and consequentially on the dynamical $\beta(t)$ measurements. 

\begin{figure}[t]
\includegraphics[width=0.95\columnwidth]{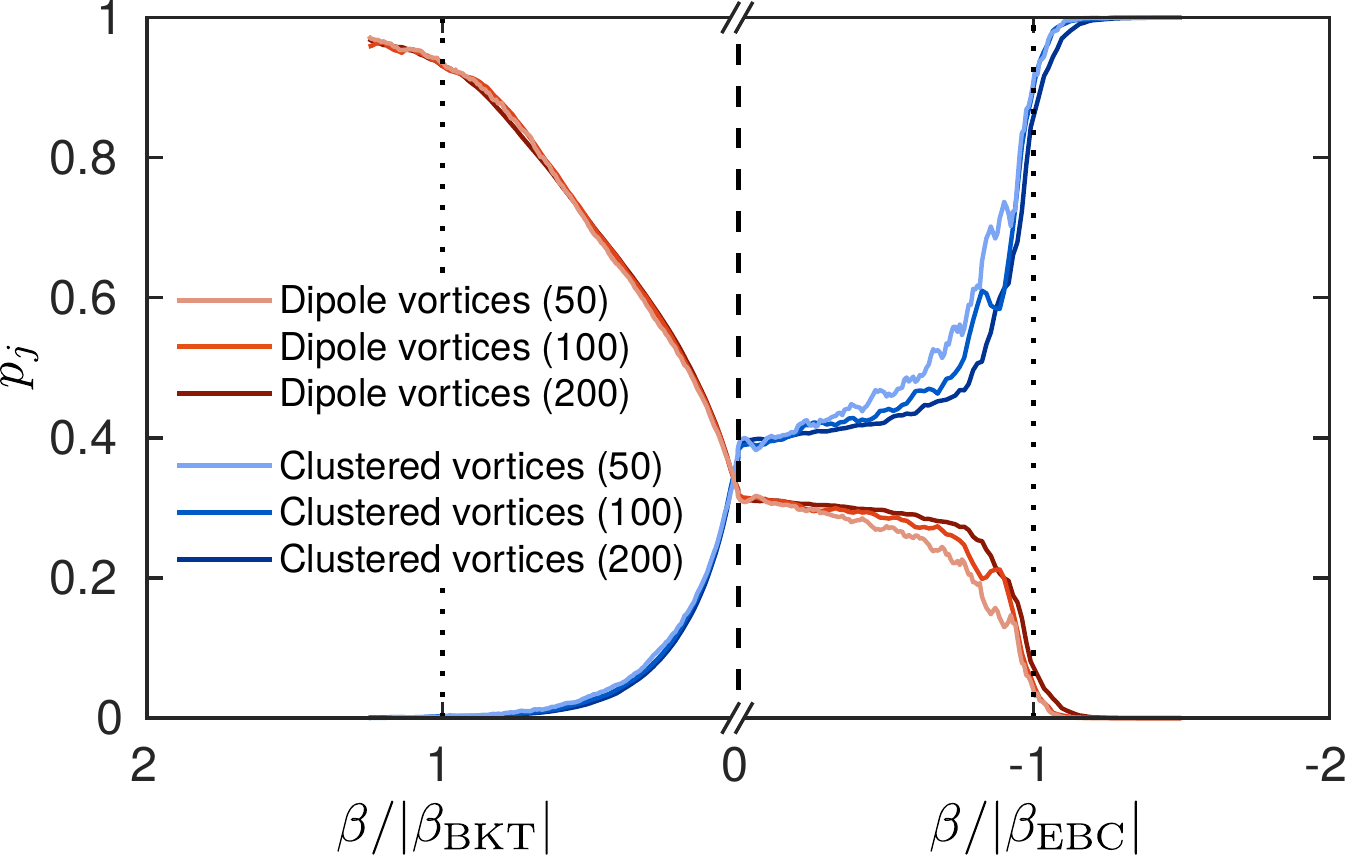}
\caption{Fractional populations $p_j=N_j/N_v$ of two components of the vortex gas (where $j \in \lbrace c$: clustered vortices, $d$: dipole vortices$\rbrace$) as functions of inverse temperature $\beta$, for three different vortex numbers (indicated in brackets in the legend). The negative temperature axis is scaled by the critical temperature $|\beta_{\rm EBC}|$, and the positive temperature axis by $|\beta_{\rm BKT}|$, causing an apparent discontinuity in the slopes at $\beta=0$. The dashed vertical line indicates $\beta=0$ and the dotted vertical lines highlight the two critical temperatures. See also Fig.~1 of the main text.}
\label{fig:S3}
\end{figure}

We have repeated our Monte Carlo simulations with $N_v = 100$ and $N_v=200$ (as argued in the main text, the vortex gas appears to be out of equilibrium for $N_v \gtrsim 200$ for case I, and hence any vortex numbers beyond $200$ are not relevant for thermometry here). The obtained $p_c(\beta)$ and $p_d(\beta)$ curves are shown in Fig.~\ref{fig:S3} for all three values of $N_v$. The thermometry curves show very little variation as $N_v$ is changed, especially in the positive temperature region. Most importantly, the curves are always monotonic, regardless of $N_v$, and hence can always be used for thermometry.

\begin{figure}[t!]
\includegraphics[width=0.95\columnwidth]{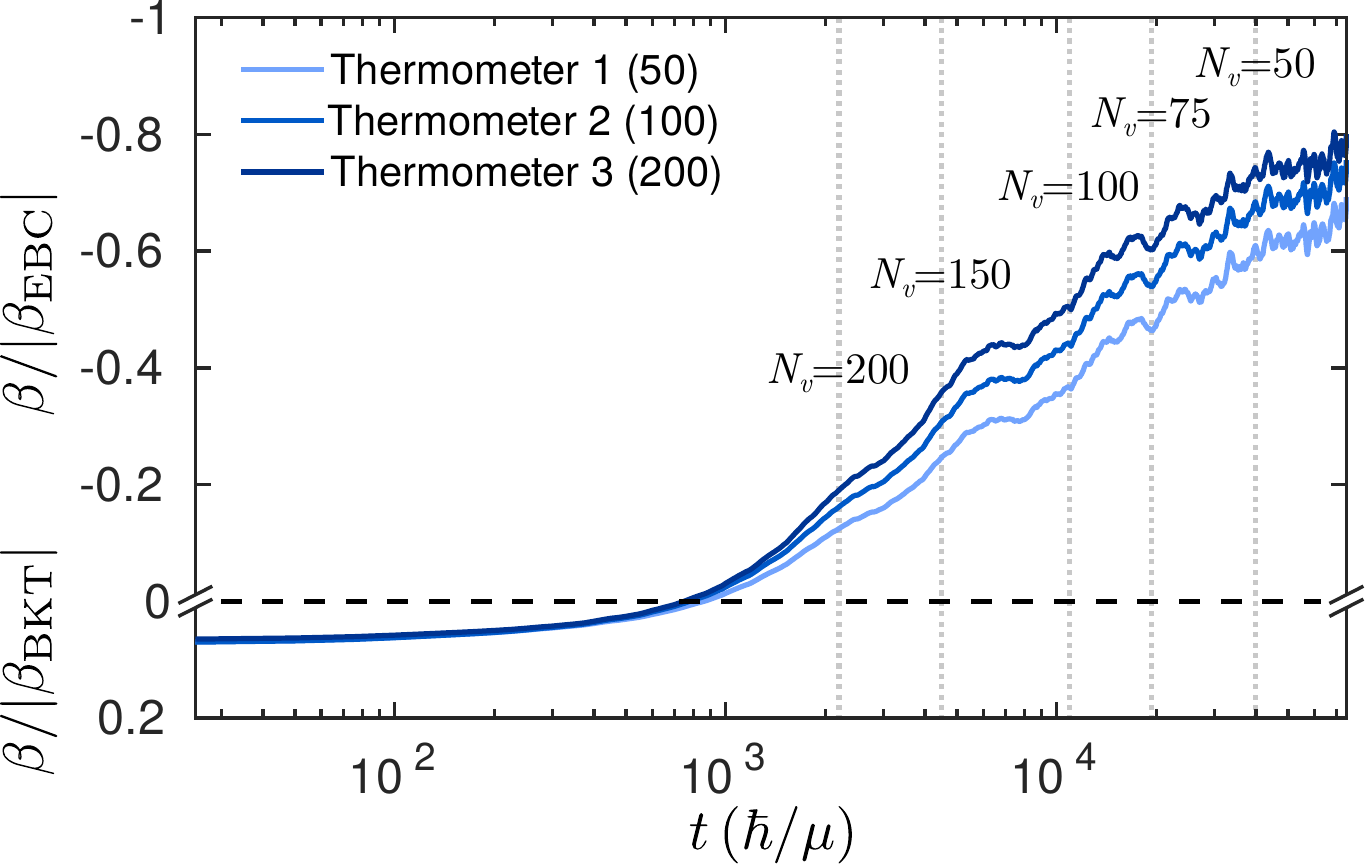}
\caption{Inverse temperature of the vortex gas as a function of time, averaged over a set of 80 dynamical GPE simulations for case I, and measured using three different thermometers, calibrated using $N_v=\lbrace 50, 100, 200\rbrace$, respectively. The temperature is measured using the populations of clusters only. As in Fig.~\ref{fig:S3}, the positive and negative temperature regions have been scaled by their respective critical temperatures, and a dashed horizontal line denotes $\beta=0$. The mean number of vortices remaining in the system is indicated at particular times by vertical dotted lines. See also Fig.~4 of the main text.}
\label{fig:S4} 
\end{figure}


Using the three cluster thermometers in Fig.~\ref{fig:S3}, we have remeasured the dynamical temperature $\beta(t)$ from our Gross--Pitaevskii simulations, and the resulting curves are presented in Fig.~\ref{fig:S4}. Evidently, our measurement using the $N_v=50$ thermometer slightly underestimates the temperature at early times ($50 \lesssim N_v \lesssim 200$), and slightly overestimates it at the latest times ($N_v\lesssim 50$). Despite this minor quantitative correction, the qualitative behaviour of vortex heating---our main conclusion from the data---is unchanged regardless of which thermometer is used. Note that we have measured these temperatures using the spline fits to the data in Fig.~\ref{fig:S3}, as described in the main text. The fluctuations that are visible in the $\beta(t)$ curves arise from the variations in $p_c(t)$, which then appear in each temperature measurement.

We have chosen to present the $N_v=50$ thermometer in the main text primarily because the vortex number is in the range of $30 \lesssim N_v \lesssim 70$ for a majority of the evolution time. Additionally, $N_v=50$ is currently an experimentally feasible vortex number for two-dimensional turbulence in a Bose--Einstein condensate, whereas systems with $N_v \gtrsim 100$ are currently challenging to produce and image.

\end{bibunit}

\end{document}